\documentstyle[12pt]{article}
\newcommand{\preprint}[1]{\begin{table}[t]  
            \begin{flushright}		    
            \begin{large}{#1}\end{large}    
            \end{flushright}		    
            \end {table}}			    

\preprint{TAUP-2272-95}
\begin{document}

\title{Distances on a one-dimensional lattice from noncommutative geometry}
\author{E. Atzmon\thanks{%
atzmon@ccsg.tau.ac.il}  \\ 
Raymond and Beverly Sackler Faculty of Exact Sciences,\\ School of Physics
and Astronomy.\\Tel\ - Aviv University. }
\date{July 2, 1995}
\maketitle

\begin{abstract}
In the following paper we continue the work of
Bimonte-Lizzi-Sparano\cite{B.L.S} on distances on a one dimensional lattice. We
succeed in proving analytically the
exact formulae for such distances. We find that the distance to an even point
on the lattice is the geometrical average of the
``predecessor'' and ``successor'' distances to the neighbouring odd points.
\end{abstract}
\newpage
\section{Distances on a one dimensional lattice from noncommutative geometry
}

The distance between two points a and b on a given space, is
defined, from the geometrical point of view as the infimum of all lengths
of paths connecting the two points on that space.

\begin{equation}
\label{eq.1}d(a,b)=\;\inf \;l_\alpha \left( a,b\right)
\end{equation}

where $l_\alpha \;$is the length of the path $\alpha \;$which connects the
points $a$\ and $b$.\ This definition is simple and well understood,
when dealing with smooth metric manifolds. However the above definition is
not applicable to discrete spaces in which the term ``path'' is not well
defined. In order to be able to define a distance on such spaces too,
one has to use tools other than the usual geometrical ones. Those tools are
found in noncommutative geometry (n.c.g). The n.c.g definition for distance
is:\cite{Connes}
\begin{equation}
\label{eq.2}d(a,b)=\sup _f\left\{ \left| f(a)-f(b)\right| \;:\;f\in
A,\;\left\| \;\left[ \;D,f\;\right] \;\right\| \leq 1\right\}
\end{equation}
where $a,b\in X,\;f\in A,$\ $A$ is the algebra of functions on $X,$\ $D$ is a
Dirac operator (which is a self adjoint operator with compact resolvent)
acting in the Hilbert space $H,$\ and the norm on the r.h.s is the norm of
operators in $H$. Both definitions give the same result when the base
space is a Riemannian manifold. However the n.c.g definition has the
advantage of being applicable to discrete spaces too.

In the following we continue the work of Ref\cite{B.L.S} (which we denote
``BLS'') by finding and proving the exact formula for distances on a
one dimensional lattice. We use the BLS notation. As a first step we
summarize briefly the BLS work relevant to a one dimensional lattice.

A one dimensional lattice is defined as:

\begin{equation}
\label{eq.3}x_k=ak,\quad k\in Z
\end{equation}

In order to compute the distance between two point $x_0$\ and $x_k$ one has
first to define the Dirac operator. BLS used the Wilson definition
for a Dirac operator:

\begin{equation}
\label{eq.4}\left( D\Psi \right) _x=\frac 1{2ia}\left( \Psi _{x+1}-\Psi
_{x-1}\right)
\end{equation}
where $D$ acts on the Hilbert space of square integrable spinors $\Psi $ $%
\;\left( \Psi \in L_2\right) $, and $x$ is a site on the one dimensional
lattice. Let $f$ act on the Hilbert space as a linear operator. It is shown
that in order to get the supremum it is enough to take the supremum over the
set of real functions:

\begin{equation}
\label{eq.5}f^{\left( k\right) }\equiv \left\{ f_i^{\left( k\right) }\quad
;\quad i\in Z\right\}
\end{equation}
such that:

\begin{equation}
\label{eq.6}
\begin{array}{l}
\mbox{a)\qquad }f_i^{\left( k\right) }=f_0^{\left( k\right) }\quad \forall
\;i<0 \\ \mbox{b)\qquad }f_i^{\left( k\right) }=f_k^{\left( k\right) }\quad
\forall \;i>k \\ \mbox{c)\qquad \ }f_i^{\left( k\right) }\leq f_j^{\left(
k\right) }\quad \forall \;i<j \\ \mbox{d)\qquad }\left\| \left[
D\;,\;f^{\left( k\right) }\right] \right\| \leq 1
\end{array}
\end{equation}
BLS then show that, following Eq.(\ref{eq.2}),(\ref{eq.6}) the distance
becomes :

\begin{equation}
\label{eq.7}d_k=\sup \left\{ f_k^{\left( k\right) }-f_0^{\left( k\right)
}\;\;:\;\left\| \left[ D\;,\;f^{\left( k\right) }\right] \right\| \leq
1\right\}
\end{equation}
where the norm on the r.h.s is equal to the maximum eigenvalue $r$\ of a
square, symmetric and real ``three-diagonal'' matrix \ H$^{\left( k\right) }$\
:

\begin{equation}
\label{eq.8}\left(
\begin{array}{ccccc}
0 & \Delta _1 & 0 & 0 & 0 \\
\Delta _1 & 0 & \Delta _2 & 0 & 0 \\
0 & \Delta _2 & 0 & \ddots & 0 \\
0 & 0 & \ddots & \ddots & \Delta _k \\
0 & 0 & 0 & \Delta _k & 0
\end{array}
\right)
\end{equation}
where $\Delta _i=f_i^{\left( k\right) }-f_{i-1}^{\left( k\right) }$\ .

Since all the elements of H$^{\left( k\right) }$ are not negative, one can
use a theorem\cite{Theorem} according to which, in such case the maximum
eigenvalue of H$%
^{\left( k\right) }$ is less than 1, if and only if, all the leading principal
minors H$_n\left( \Delta _1,\ldots ,\Delta _{n-1}\right) $ where $n=1,\ldots
,k+1$ of the matrix \ I$\;-\;$H are positive\ :

\begin{equation}
\label{eq.9}\mbox{H}_1^{\left( k\right) }=1>0,\ldots ,\mbox{H}_i^{\left(
k\right) }>0,\ldots ,\mbox{H}_k^{\left( k\right) }>0,\;\mbox{H}%
_{k+1}^{\left( k\right) }=\det \left( \mbox{I}^{\left( k\right) }\;-\;\mbox{H%
}^{\left( k\right) }\right) =0
\end{equation}

It is then shown that in order to find the distance, i.e. the supremum of $%
\sum_i\triangle _i$ satisfying all conditions set by (\ref{eq.9}), one can
use the method of Lagrange multipliers, and solve the following set of
equations in the unknowns $\left( \triangle _1^{\left( k+1\right) }\ldots
\triangle _k^{\left( k+1\right) }\right) \;$and $\alpha \;$(the Lagrange
multiplier)\ :

\begin{equation}
\label{eq.10}\left\{
\begin{array}{l}
\frac \partial {\partial \triangle _j}\left[ \sum_{i=1}^k\triangle _i+\alpha
\mbox{H}_{k+1}^{\left( k\right) }\right] _{\triangle _j=\triangle _j^{\left(
k+1\right) }}=0,\qquad \forall \;\;j\leq k \\ \mbox{H}_{k+1}^{\left(
k\right) }=0
\end{array}
\right\}
\end{equation}
to find what the distance is.

In the following we solve exactly the above equations.

\section{Exact solution}

As the first step in solving the above equations we have to use some of the
properties of real, symmetric, three-diagonal matrices:\

The first property is :

\begin{equation}
\label{eq.11}\mbox{H}_{k+1}=\mbox{H}_k-\triangle _k^2\mbox{H}_{k-1}
\end{equation}
from which it follows that:
\begin{equation}
\label{eq.12}
\begin{array}{l}
\mbox{a)\ \ \ }\triangle _k=\left( \frac{\mbox{H}_k-\mbox{H}_{k+1}}{\mbox{H}%
_{k-1}}\right) ^{\frac 12}\mbox{\qquad if \ H}_{k-1}\neq 0 \\  \\
\mbox{b)\ \ \ \ }\frac \partial {\partial \triangle _k}\mbox{H}%
_{k+1}=-2\triangle _k\mbox{H}_{k-1}\qquad \forall \;k \\ \mbox{(because
there is no\ }\triangle _k\mbox{ \ in \ H}_k\mbox{)} \\  \\
\mbox{c)\ \ \ \ }\frac \partial {\partial \triangle _{k-1}}\mbox{H}%
_{k+1}=\frac \partial {\partial \triangle _{k-1}}\mbox{H}_k\qquad \forall
\;k \\ \mbox{(because there is no\ }\triangle _{k-1}\mbox{\ \ in \ H}_{k-1}%
\mbox{)}
\end{array}
\end{equation}

We now make use of the fact that\ :
\begin{equation}
\label{eq.13}\frac \partial {\partial x}\det
A_{k+1}=\sum_{i=1}^{k+1}A_{i,k+1}^{\left( x\right) }
\end{equation}
where $A_{k+1}$\ is a $\left( k+1\right) \times \left( k+1\right) $\ matrix
and :
\begin{equation}
\label{eq.14}A_{i,k+1}^{\left( x\right) }\equiv \det \left(
\begin{array}{l}
\left( \;\;\;\;\;\;\;\vdots \;\;\;\;\;\;\right) _{1\leq j\leq i-1} \\
\left( \frac \partial {\partial x},\ldots ,\frac \partial {\partial
x}\right) _{j=i} \\
\left( \;\;\;\;\;\;\;\vdots \;\;\;\;\;\;\right) _{k+1\geq j\geq i+1}
\end{array}
\right)
\end{equation}
i.e. a determinant of the matrix where the derivation is imposed on the $\;i$
$-th$\ \ row.

However since \ I $-$ H\ \ is also a three-diagonal matrix, $\;\triangle
_j$\ \ is found only in the $\;j-th$\ \ and $\;(j+1)-th$\ \ rows. One thus
gets\ :
\begin{equation}
\label{eq.15}\frac \partial {\partial \triangle _j}\mbox{H}_{k+1}=\mbox{H}%
_{j,k+1}^{\left( \triangle _j\right) }+\mbox{H}_{j+1,k+1}^{\left( \triangle
_j\right) }
\end{equation}

One can show (using an even number of inter-changes between rows and columns)
that: H$_{j,k+1}^{\left( \triangle _j\right) }=$ \ H$_{j+1,k+1}^{\left(
\triangle _j\right) }$, yielding,

\begin{equation}
\label{eq.16}\frac \partial {\partial \triangle _j}\mbox{H}%
_{k+1}=-2\triangle _j\mbox{H}_{j,k+1}^{\left( \triangle _j\right)
}=-2\triangle _{j}\mbox{H}_{j+1,k+1}^{\left( \triangle _j\right) }
\end{equation}

And by using the block structure of \ H$_{j,k+1}^{\left( \triangle _j\right)
}$\ \ one gets that :

\begin{equation}
\label{eq.17}\frac \partial {\partial \triangle _j}\mbox{H}%
_{k+1}=-2\triangle _j\mbox{H}_{j-1}\left( \triangle _{1,}\ldots ,\triangle
_{j-2}\right) \cdot \mbox{H}_{k-j}\left( \triangle _i\rightarrow \triangle
_{k-i+1}\;:\;\forall \;i\leq k-j-1\right)
\end{equation}
but since it was shown\cite{B.L.S} that the maximum is unique, and since the
equations (\ref{eq.10}) have a symmetry under \ $\triangle
_i\longleftrightarrow \triangle _{k-i+1}$ \, the solution must fix $\;\triangle
_i=\triangle _{k-i+1}$. Thus what one essentially gets is
simply that:
\begin{equation}
\label{eq.18}\frac \partial {\partial \triangle _j}\mbox{H}%
_{k+1}=-2\triangle _j\mbox{H}_{j-1}\mbox{H}_{k-j}
\end{equation}
Equations (\ref{eq.10}) take the following form:
\begin{equation}
\label{eq.19}
\begin{array}{c}
\left\{
\begin{array}{c}
1-2\alpha \triangle _j
\mbox{H}_{j-1}\mbox{H}_{k-j}=0 \\ \vdots
\end{array}
\right\} _{1\leq j\leq k} \\
\mbox{H}_{k+1}=0
\end{array}
\end{equation}

It can be verified, that if \ H$_{k-1}$\ \ and \ H$_k$\ \ are both
$\;\neq 0$, it follows from the last equation that \ $\triangle
_k=\left( \frac{\mbox{H}_k}{\mbox{H}_{k-1}}\right) ^{\frac 12}$; \ while from
the \ $j=k\;\;$equation one gets, $\;\alpha =\frac 1{2\sqrt{\mbox{H}_k%
\mbox{H}_{k-1}}}$\ \ .

Until now what we have done is just to simplify the equations that we have
to solve.\ From now on we assume what the solutions for the unknowns should
be. Since the solution is unique, then by setting the assumed solutions\
(i.e. the \ $\triangle _i\;\;$'s)\ into the set of equations (\ref{eq.19}%
) and showing that they really solve the equations,\ we are essentially
proving that the assumed solutions are correct.\ Knowing the
solutions for the \ $\triangle _i$\ 's one can then find what the distance
is.\ As the first case we will solve for $k=$ even.

\subsection{The $\;k=even$\ \ case}

Let us assume that the solution for the $\;\triangle _i\;$'s\ is:
\begin{equation}
\label{eq.20}\triangle _i^{\left( k\right) }=\frac{\frac 12\left( 1-\left(
-1\right) ^i\right) \left( \frac k2+1\right) +\left( -1\right) ^i\left[
\frac{i+1}2\right] }{\sqrt{\frac k2\left( \frac k2+1\right) }}%
\;\;\;\;\forall \;i\leq k
\end{equation}
where the\ $\left[ \;\right] \;\;$bracket stand for the integral value of
the term within;\ and the Lagrange multiplier is :\
\begin{equation}
\label{eq.21}\alpha ^{\left( k\right) }=\frac{\left[ k\left( k+2\right)
\right] ^{\frac{k-1}2}}{2^{k-1}k\left[ \left( \frac k2-1\right) !\right] ^2}
\end{equation}

By using the recursion relation (\ref{eq.11}), and from the fact
that \ H$_0=$H$_1=1\;\;$, one can prove by induction that:
\begin{equation}
\label{eq.22}\mbox{H}_{2l+1}^{(k)}=\frac{2^{2l}l!\left( \frac k2-1\right) !}{%
\left[ k\left( k+2\right) \right] ^l\left( \frac k2-l-1\right) !}%
\;\;\;\;\;:\;1\leq l\leq \frac k2\;\;\;\;\;\;\left( \Rightarrow \;\mbox{H}%
_{k+1}=0\right)
\end{equation}
and

\begin{equation}
\label{eq.23}\mbox{H}_{2l}^{\left( k\right) }=\frac{2^{2l}l!\left( \frac
k2\right) !}{\left[ k\left( k+2\right) \right] ^l\left( \frac k2-l\right) !}%
\;\;\;\;\;\;\;:\;1\leq l\leq \frac k2
\end{equation}

We now have all the ingredients we need for the equations. All we have
to do is to check that essentially all the equations are fulfilled; and
indeed they are, as can be verified by a few arithmetical steps.

After proving that the\ $\triangle _i\;$'s \ are the solution for the set of
equations (\ref{eq.19}) we can now find exactly, that the distance in
the\ $k=even\;\;$case is\ :
\begin{equation}
\label{eq.24}d_k=2a\cdot \sum_{i=1}^k\triangle _i^{\left( k\right) }=2a\cdot
\sqrt{\frac k2\left( \frac k2+1\right) }=a\sqrt{k\left( k+2\right) }
\end{equation}
Q.E.D. for the $k=even$ case.

\subsection{The\ $k=odd$\ \ case\ :}

Let us assume that the solution in this case has the following structure:

\begin{equation}
\label{eq.27}
\begin{array}{l}
\triangle _{2i+1}^{\left( k\right) }=\left( 1-\varepsilon \right) \\
\\
\triangle _{2i}^{\left( k\right) }=\varepsilon \\
\\
\alpha ^{\left( k\right) }=\frac 1{2\varepsilon ^{\frac{k-1}2}\left(
1-\varepsilon \right) ^{\frac{k+1}2}}
\end{array}
\end{equation}
and let us also assume temporarily that \ H$_1^{\left( k\right)
}=1-\varepsilon $. It is easy to show by the recursion relation
(\ref{eq.11}) that:
\begin{equation}
\label{eq.28}
\begin{array}{l}
\mbox{H}_{2j}^{\left( k\right) }=\varepsilon ^j\left( 1-\varepsilon \right)
^j \\  \\
\mbox{H}_{2j+1}^{\left( k\right) }=\varepsilon ^j\left( 1-\varepsilon
\right) ^{j+1}
\end{array}
\;\forall \;0\leq j\leq \frac{k-1}2
\end{equation}
The solutions can now be set into the equations (\ref{eq.19}). It is easy
to verify that those solutions obey the set
of equations.\ However since we know that \ H$_1^{\left( k\right) }=1$ \
(rather then $1-\varepsilon $)\ \ one has to take the limit where\ $%
\varepsilon \rightarrow 0^{+}$.\ \ Thus, when the limit $\;\varepsilon
\rightarrow 0^{+}$\ \ is applied, it follows that:
\begin{equation}
\label{eq.26}\mbox{H}_0^{\left( k\right) }=\mbox{H}_1^{\left( k\right)
}=1,\;\;\;\mbox{H}_j^{\left( k\right) }=0\;\;\forall \;j\geq 2
\end{equation}
which means that all the conditions in Eq.(\ref{eq.9}) are fulfilled. Taking
the limit\ $\varepsilon \rightarrow 0^{+}$\ \ it follows that:
\begin{equation}
\label{eq.25}
\begin{array}{c}
\triangle _{2j+1}=1 \\
\\
\triangle _{2j}=0 \\
\\
\alpha \rightarrow \infty
\end{array}
\;\forall \;0\leq j\leq \frac{k-1}2
\end{equation}

We were able to use the limiting procedure, through the fact that this is a
supremum problem - so we just had to prove that the limit exists. By that
we have essentially proven that the solutions are as listed
in Eq.(\ref{eq.25}).

Having all the $\triangle _i^{\left( k\right) }\;$'s\ we can find that the
distance in the $k=odd$\ case is:
\begin{equation}
\label{eq.29}d_k=2a\sum_{i=1}^k\triangle _i^{\left( k\right) }=a\left(
k+1\right)
\end{equation}
Q.E.D. for the $k=odd$\ case.

We can summarize that the distances in the lattice are:

\begin{equation}
\label{eq.30}\left\{
\begin{array}{l}
d_{2j-1}=2aj \\
\\
d_{2j}=2a\sqrt{j\left( j+1\right) }
\end{array}
\right\}
\end{equation}

\section{Discussion}

\subsection{Mathematical aspects}

All distances have anomalous behavior, as compared to
what one expects classically. The anomaly in the odd point distances is
exactly equal to one. The anomaly in the even point distances depends on the
points. These are non-constant, irrational numbers, smaller than
one, which asymptotically tend to 1.

The second outcome from (\ref{eq.30}) is that the distance to an even
point is the geometrical average of the distances to the nearby predecessor
and successor odd points (rather than an arithmetical average which one
would have expected classically).

Concerning the large $k$ \ limit, it follows where $k=2j\rightarrow \infty $
\ the distance to an even point $k=2j$ \ has the following behavior:
\begin{equation}
\label{eq.31}2a\sqrt{j\left( j+1\right) }\longrightarrow \;\;2aj\left(
1+\frac 1{2j}\right)
\end{equation}
One can thus see that the asymptotic behavior of the even point distances
becomes the same as for the odd point distances (in other words
the geometrical average asymptotically tends to the arithmetical average).

\subsection{Physical aspects}

One can say that the cause for the results we got to be different from what
one might expect, is an outcome of using the local discrete Wilson\ - Dirac
operator. One should perhaps try non-local Dirac operator in order to
get the classical behavior that we expect.

However as far as the physical world is concerned, we have shown that
asymptotically, in the large $k$ limit, the differnce in behavior of the odd
point distances and
the even point distances on the lattice vanishes.\ Usually in
nature one deals with lattices with very large $k$. Thus all these effects
are not seen. One could treat the lattice as a ''quantum''
system, in which the quantum behavior of nature is revealed for small $k$%
, and for large $k$ the classical behavior is reached asymptotically\ (We
remind the reader that the n.c.g definition for distance considers not only the
base
space, but also the Hilbert space). Thus, following this point of view, one
can continue working with the local Wilson - Dirac operator, though revealing
non-classical behavior of the lattice.

Acknowledgments

I would like to thank Prof. Y. Ne'eman, Prof. F. Lizzi and Mr. O. Kennet for
useful discussions.

\end{document}